\documentclass{ws-procs11x85}

\def\sPlots{\hbox{$_s$}{\cal P}lots}
\def\sPlot{\hbox{$_s$}{\cal P}lot}
\def\eLik{{\cal L}}
\def\Ns{{\rm N}_{\rm s}}
\def\f{{\rm f}}
\def\Mtrue{{\hbox{\bf{M}}}}
\def\nast{{\rm n}}
\def\inPlot{_{\rm in}{\cal P}lot}
\def\inPlots{_{\rm in}{\cal P}lots}
\def\M{{\rm M}}
\def\Deltax{{\delta x}}
\def\chiant{e \subset \Deltax}
\def\V{\hbox{\bf V}}
\def\sP{{_s{\cal P}}}

\newcommand{\deltaE}{\Delta E}

\def\P{{\cal P}}
\def\d{{\rm d}}
\def\Bz{B^0}
\def\to{{\rightarrow}}
\def\pipi{\pi^+\pi^-}

\def\kppim{K^+\pi^-}
\def\epem{e^+e^-}
\def\qqbar{q\overline q}
\def\NIM{{\em Nucl. Inst. Meth.}}
\def\id{{\em i.e.}}

\usepackage{multicol}
\def\Journal#1#2#3#4{{#1} {\bf #2}, #3 (#4)}

\makeindex
\begin{document}

\title{{\LARGE $\sPlot:$} A QUICK INTRODUCTION}

\author{M.~PIVK}

\address{CERN, 1211 Geneva 23, Switzerland\\E-mail: muriel.pivk@cern.ch}


\maketitle\abstract{The paper advocates the use of a statistical tool dedicated to the exploration of data samples populated by several sources of events. This new technique, called $\sPlot$, is able to unfold the contributions of the different sources to the distribution of a data sample in a given variable. The $\sPlot$ tool applies in the context of a Likelihood fit which is performed on the data sample to determine the yields of the various sources.}

\begin{multicols}{2}

\baselineskip=13.07pt

\section{Introduction}
This paper describes a new technique to explore a data sample when the latter consists of several sources of events merged into a single sample of events. The events are assumed to be characterized by a set of variables which can be split into two components. The first component is a set of variables for which the distributions of all the sources of events are known: below, these variables are referred to as the {\em discriminating} variable. The second component is a set of variables for which the distributions of some sources of events are either truly unknown or considered as such: below, these variables are referred to as the {\em control} variables.

The new technique, termed $\sPlot$~\footnote{The $\sPlot$ technique is the subject of a publication~\cite{bib:sPlot} where details of the calculations and more examples can be found.}, allows one to reconstruct the distributions for the control variable, independently for each of the various sources of events, without making use of any {\em a priori} knowledge on this variable. The aim is thus to use the knowledge available for the discriminating variables to be able to infer the behavior of the individual sources of events with respect to the control variable. An essential assumption for the $\sPlot$ technique to apply is that the control variable is uncorrelated with the discriminating variables.
 
The $\sPlot$ technique is developed in the context of a maximum Likelihood method making use of the discriminating variables. Section~\ref{sec:basics} is dedicated to the definition of fundamental objects necessary for the following. Section~\ref{sec:inPlot} presents an intermediate technique, simpler but inadequate, which is a first step towards the $\sPlot$ technique. The $\sPlot$ formalism is then developed Section~\ref{sec:sPlot} and its properties explained in Section~\ref{sec:sPlotprop}. An example of $\sPlot$ at work is provided in Section~\ref{sec:Illustrations} and some applications are described in Section~\ref{sec:app}. Finally, the case where the control variable is correlated with the discriminating ones is discussed in Section~\ref{sec:corr}.

\section{Basics and definitions}
\label{sec:basics}
One considers an unbinned extended maximum Likelihood analysis of a data sample in which are merged several species (signal and background) of events. The log-Likelihood is expressed as:
\begin{equation}
\label{eq:eLik}
\eLik=\sum_{e=1}^{N}\ln \Big\{ \sum_{i=1}^{\Ns}N_i\f_i(y_e) \Big\} -\sum_{i=1}^{\Ns}N_i ~,
\end{equation}
where 
\begin{itemize}
\item $N$ is the total number of events considered,
\item $\Ns$ is the number of species of events populating the data sample,
\item $N_i$ is the (non-integral) number of events expected on the average for the $i^{\rm th}$ species,
\item $y$ represents the set of discriminating variables, which can be correlated with each other,
\item $\f_i(y_e)$ is the value of the Probability Density Function (pdf) of $y$ for the $i^{\rm th}$ species and for event $e$.
\end{itemize}
The log-Likelihood $\eLik$ is a function of the $\Ns$ yields $N_i$ and, possibly, of implicit free parameters designed to tune the pdfs on the data sample. These parameters as well as the yields $N_i$ are determined by maximizing the above log-Likelihood.

The crucial point for the reliability of such an analysis is to use an exhaustive list of sources of events combined with an accurate description of all the pdfs~$\f_i$. If the distributions of the control variables are known (resp. unknown) for a particular source of events, one would like to compare the expected distribution for this source to the one extracted from the data sample (resp. determine the distribution for this source)~\footnote{Removing one of the discriminating variables from the set~$y$ before performing again the maximum Likelihood fit, one can consider the removed variable as a control variable~$x$, provided it is uncorrelated with the others.}.

The control variable $x$ which, by definition, does not explicitly appear in the expression of $\eLik$, can be:
\begin{enumerate}
\item totally correlated with the discriminating variables $y$ ($x$ belongs to the set $y$ for example). This is the case treated in Section~\ref{sec:inPlot}.
\item uncorrelated with $y$. This is the subject of Section~\ref{sec:sPlot}.
\item partly correlated with $y$. This case is discussed Section~\ref{sec:corr}.
\end{enumerate}
In an attempt to have access to the distributions of control variables, a common method consists of applying cuts which are designed to enhance the contributions to the data sample of particular sources of events. However, the result is frequently unsatisfactory: firstly because it can be used only if the signal has prominent features to be distinguished from the background, and secondly because of the cuts applied, a sizeable fraction of signal events can be lost, while a large fraction of background events may remain.

The aim of the $\sPlot$ formalism developed in this paper is to unfold the true distribution (denoted in boldface $\Mtrue_\nast(x)$) of a control variable~$x$ for events of the $\nast^{\rm th}$ species (any one of the $\Ns$ species), from the sole knowledge of the pdfs of the discriminating variables $\f_i$, the first step being to proceed to the maximum Likelihood fit to extract the yields $N_i$. The statistical technique $\sPlot$ allows to build histograms in $x$ keeping all signal events while getting rid of all background events, and keeping track of the statistical uncertainties per bin in $x$.

\section{First step towards $\sPlot$: $\inPlot$} 
\label{sec:inPlot}
In this Section, as a means of introduction, one considers a variable~$x$ assumed to be totally correlated with~$y$: $x$ is a function of $y$. A fit having been performed to determine the yields $N_i$ for all species, one can define naively, for all events, the weight
\begin{equation}
\label{eq:weightxiny}
\P_\nast(y_e)={N_\nast\f_\nast(y_e)\over\sum_{k=1}^{\Ns}N_k\f_k(y_e) } ~,
\end{equation}
which can be used to build an estimate, denoted $\tilde\M_\nast$, of the $x$-distribution of the species labelled $\nast$ (signal or background):
\begin{equation}
\label{eq:inPlots}
N_\nast\tilde\M_\nast(\bar x)\Deltax~\equiv ~\sum_{\chiant} \P_\nast(y_e) ~,
\end{equation}
where the sum runs over the events for which the~$x$ value lies in the bin centered on $\bar x$ and of total width~$\Deltax$. 

In other words, $N_\nast\tilde\M_\nast(\bar x)\Deltax$ is the $x$-distribution obtained by histogramming events, using the weight of Eq.~(\ref{eq:weightxiny}). To obtain the expectation value of $\tilde\M_\nast$, one should replace the sum in Eq.~(\ref{eq:inPlots}) by the integral
\begin{equation}
\label{eq:sumGOTOintegral}
\left<\sum_{\chiant}\right>
\longrightarrow
\int\d y\sum_{j=1}^{\Ns}N_j\f_j(y)\delta(x(y)-\bar x)\Deltax ~.
\end{equation}
Similarly, identifying the number of events $N_i$ as determined by the fit to the expected number of events, one readily obtains:
\begin{eqnarray}
\label{eq:naiveweightworks}
\left<
N_\nast\tilde\M_\nast(\bar x)\right>
&\equiv&N_\nast\Mtrue_\nast(\bar x) ~.
\end{eqnarray}
Therefore, the sum over events of the naive weight $\P_\nast$ reproduces, on average, the true distribution $\Mtrue_\nast(x)$. Plots obtained that way are referred to as $\inPlots$: they provide a correct means to reconstruct $\Mtrue_\nast(x)$ only insofar as the variable considered is {\bf in} the set of discriminating variables~$y$. These $\inPlots$ suffer from a major drawback:~$x$ being fully correlated to $y$, the pdfs of~$x$ enter implicitly in the definition of the naive weight, and as a result, the $\tilde\M_\nast$ distributions cannot be used easily to assess the quality of the fit, because these distributions are biased in a way difficult to grasp, when the pdfs $\f_i(y)$ are not accurate. For example, let us consider a situation where, in the data sample, some events from the $\nast^{\rm th}$ species show up far in the tail of the $\M_\nast(x)$ distribution which is implicitly used in the fit. The presence of such events implies that the true distribution $\Mtrue_\nast(x)$ must exhibit a tail which is not accounted for by $\M_\nast(x)$. These events would enter in the reconstructed $\inPlot$ $\tilde\M_\nast$ with a very small weight, and they would thus escape detection by the above procedure: $\tilde\M_\nast$ would be close to $\M_\nast$, the distribution assumed for~$x$. Only a mismatch in the core of the $x$-distribution can be revealed with $\inPlots$. Stated differently, the error bars which can be attached to each individual bin of $\tilde\M_\nast$ cannot account for the systematical bias inherent to the $\inPlots$.

\section{The $\sPlot$ formalism} 
\label{sec:sPlot}
In this Section one considers the more interesting case where the two sets of variables~$x$ and~$y$ are uncorrelated. Hence, the total pdfs~$\f_i(x,y)$ all factorize into products $\Mtrue_i(x)\f_i(y)$. While performing the fit, which relies only on $y$, no {\it a priori} knowledge of the $x$-distributions is used.
 
One may still consider the above distribution $\tilde\M_\nast$ (Eq.~(\ref{eq:inPlots})), using the naive weight of Eq.~(\ref{eq:weightxiny}). However in that case, the expectation value of $\tilde\M_\nast$ is a biased estimator of $\Mtrue_\nast$: 
\begin{eqnarray}
\left< 
N_\nast\tilde\M_\nast(\bar x) \right> &=&\int\d y\d x\sum_{j=1}^{\Ns}N_j\Mtrue_j(x)\f_j(y)\delta(x-\bar x)\P_\nast
\nonumber \\
\label{eq:sZut}
=&N_\nast& \sum_{j=1}^{\Ns}\Mtrue_j(\bar x)
N_j\int\d y{\f_\nast(y)\f_j(y)\over\sum_{k=1}^{\Ns}N_k\f_k(y)}
\\
\neq &N_\nast&\Mtrue_\nast(\bar x) ~.
\nonumber
\end{eqnarray}
Here, the naive weight is no longer satisfactory because, when summing over the events, the $x$-pdfs~$\Mtrue_j(x)$ appear now on the right hand side of Eq.~(\ref{eq:sumGOTOintegral}), while they are absent in the weight. However, one observes that the correction term in the right hand side of Eq.~(\ref{eq:sZut}) is related to the inverse of the covariance matrix, given by the second derivatives of $-\eLik$:
\begin{equation}
\label{eq:VarianceMatrixDirect}
\V^{-1}_{\nast j}~=~
{\partial^2(-\eLik)\over\partial N_\nast\partial N_j}~=~
\sum_{e=1}^N {\f_\nast(y_e)\f_j(y_e)\over(\sum_{k=1}^{\Ns}N_k\f_k(y_e))^2} ~.
\end{equation}
On average, one gets:
\begin{eqnarray}
\label{VarianceMatrixAsymptotic}
\left<
\V^{-1}_{\nast j}
\right>
&=&
\label{VarianceMatrix}
\int\d y {\f_\nast(y)\f_j(y)\over\sum_{k=1}^{\Ns}N_k\f_k(y)} ~.
\end{eqnarray}
Therefore, Eq.~(\ref{eq:sZut}) can be rewritten:
\begin{equation}
\label{eq:resZut}
\left<
\tilde\M_\nast(\bar x)
\right>
=
\sum_{j=1}^{\Ns}\Mtrue_j(\bar x)
N_j
\left<
\V^{-1}_{\nast j}
\right>
 ~.
\end{equation}
Inverting this matrix equation, one recovers the distribution of interest:
\begin{equation}
N_\nast \Mtrue_\nast(\bar x)=\sum_{j=1}^{\Ns} 
\left<
\V_{\nast j}
\right>
\left<
\tilde\M_j(\bar x)
\right>
 ~.
\end{equation}
Hence, when~$x$ is uncorrelated with the set $y$, the appropriate weight is not given by Eq.~(\ref{eq:weightxiny}), but is the covariance-weighted quantity (thereafter called sWeight) defined by:
\begin{equation}
\begin{Large}
\label{eq:weightxnotiny}
\fbox{
$ \sP_\nast(y_e)={\sum_{j=1}^{\Ns} \V_{\nast j}\f_j(y_e)\over\sum_{k=1}^{\Ns}N_k\f_k(y_e) } $
}
\end{Large} ~.
\end{equation}
With this sWeight, the distribution of the control variable~$x$ can be obtained from the $\sPlot$ histogram:
\begin{equation}
\label{eq:masterequation}
{N_\nast}\ _s\tilde\M_\nast(\bar x)\Deltax ~\equiv~ \sum_{\chiant}  \sP_\nast(y_e) ~,
\end{equation}
which reproduces, on average, the true binned distribution:
\begin{equation}
\label{eq:sPlotsFormula}
\left<
{N_\nast}\ _s\tilde\M_\nast(x)
\right>
~=~ N_\nast \Mtrue_\nast(x) ~.
\end{equation}
The fact that the covariance matrix $\V_{ij}$ enters in the definition of the sWeights is enlightening: in particular, the sWeight can be positive or negative, and the estimators of the true pdfs are not constrained to be strictly positive.

\section{$\sPlot$ properties}
\label{sec:sPlotprop}
Beside satisfying the essential asymptotic property Eq.~(\ref{eq:sPlotsFormula}), $\sPlots$ bear properties which hold for finite statistics.

The distribution $_s\tilde\M_\nast$ defined by Eq.~(\ref{eq:masterequation}) is guaranteed to be normalized to unity and the sum over the species of the $\sPlots$ reproduces the data sample distribution of the control variable. These properties rely on maximizing the Likelihood:
\begin{itemize}
\item{}
Each $x$-distribution is properly normalized. The sum over the $x$-bins of $N_\nast\ _s\tilde\M_\nast\Deltax$ is equal to $N_\nast$:
\begin{equation}
\label{eq:NormalizationOK}
\sum_{e=1}^{N} \sP_{\nast}(y_e) ~=~N_\nast ~.
\end{equation}
\item{}
In each bin, the sum over all species of the expected numbers of events equals to the number of events actually observed. In effect, for any event:
\begin{equation}
\label{eq:numberconservation}
\sum_{l=1}^{\Ns} \sP_l(y_e) ~=1 ~.
\end{equation}
\end{itemize}
Therefore, an $\sPlot$ provides a consistent representation of how all events from the various species are distributed in the control variable~$x$. Summing up the $\Ns$ $\sPlots$, one recovers the data sample distribution in~$x$,
and summing up the number of events entering in a $\sPlot$ for a given species,
one recovers the yield of the species, 
as it is provided by the fit. For instance, if one observes an excess of events for a particular $\nast^{\rm th}$ species, in a given $x$-bin, this excess is effectively accounted for in the number of events $N_\nast$ resulting from the fit. To remove these events implies a corresponding decrease in $N_\nast$. It remains to gauge how significant is an anomaly in the $x$-distribution of the $\nast^{\rm th}$ species.

The statistical uncertainty on ${N_\nast}\  _s\tilde\M_\nast(x) \Deltax$ can be defined in each bin by 
\begin{equation}
\label{eq:ErrorPerBin}
\sigma[N_\nast\  _s\tilde\M_\nast(x) \Deltax]~=~\sqrt{\sum_{\chiant} (\sP_\nast)^2} ~.
\end{equation}
The above properties~Eqs.~(\ref{eq:sPlotsFormula})-(\ref{eq:numberconservation}) are completed by the fact that the sum in quadrature of the uncertainties Eq.~(\ref{eq:ErrorPerBin}) reproduces the statistical uncertainty on the yield $N_\nast$, as it is provided by the fit. In effect, the sum over the $x$-bins reads:
\begin{equation}
\label{eq:SumOfErrors}
\sum_{[\Deltax]}\sigma^2[N_\nast\  _s\tilde\M_\nast \Deltax]
~=~\V_{\nast \nast}~.
\end{equation}
Therefore, for the expected number of events per $x$-bin indicated by the $\sPlots$, 
the statistical uncertainties are straightforward to compute using Eq.~(\ref{eq:ErrorPerBin}).
The latter expression is asymptotically correct, and it provides a consistent representation of how the overall uncertainty on $N_\nast$ is distributed in~$x$ among the events of the $\nast^{\rm th}$ species. Because of Eq.~(\ref{eq:SumOfErrors}), and since the determination of the yields is optimal when obtained using a Likelihood fit, one can conclude that the $\sPlot$ technique is itself an optimal method to reconstruct distributions of control variables.

\section{Illustrations}
\label{sec:Illustrations}
An example of $\sPlot$ at work is taken from the analysis where the method was first used~\cite{bib:TheseMu,bib:pipi2002}. One deals with a data sample in which three species are present: $\Bz \to \pipi$ and $\Bz \to \kppim$ are signals and the main background comes from $\epem \to \qqbar$. The variable which is not incorporated in the fit is called $\deltaE$ and is used here as the control variable~$x$. The detailed description of the variables can be found in Refs.~\cite{bib:TheseMu,bib:pipi2002}.

The left plot of Fig.~\ref{fig:dE} shows the distribution of $\deltaE$ after applying a cut on the Likelihood ratio. Therefore, the resulting data distribution concerns a reduced subsample for which statistical fluctuations cannot be attributed unambiguously to signal or to background. For example, the excess of events appearing on the left of the peak is likely to be attributed to a harmless background fluctuation. 

\begin{figurehere}
\center
\psfig{figure=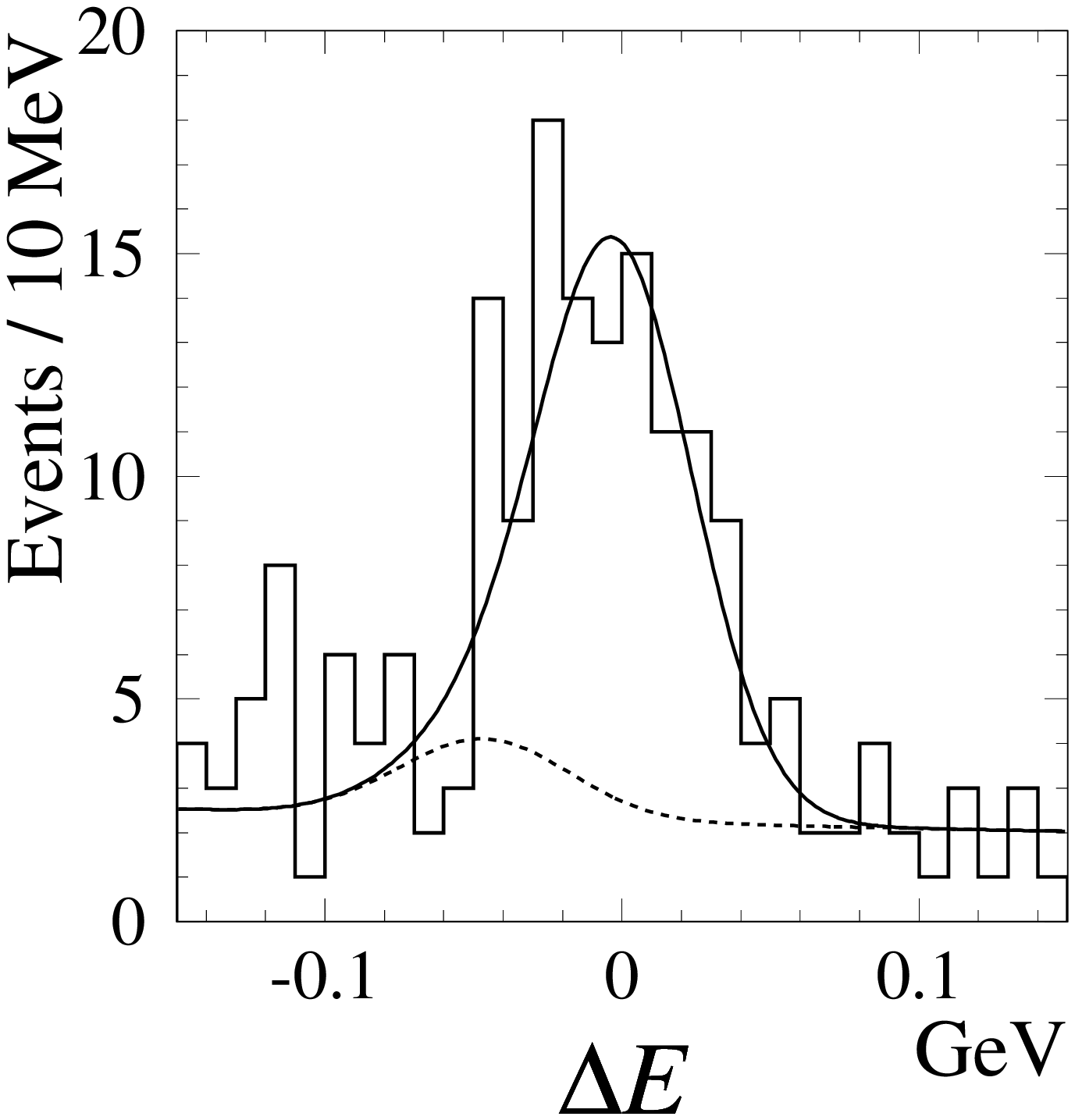,height=1.55in}
\psfig{figure=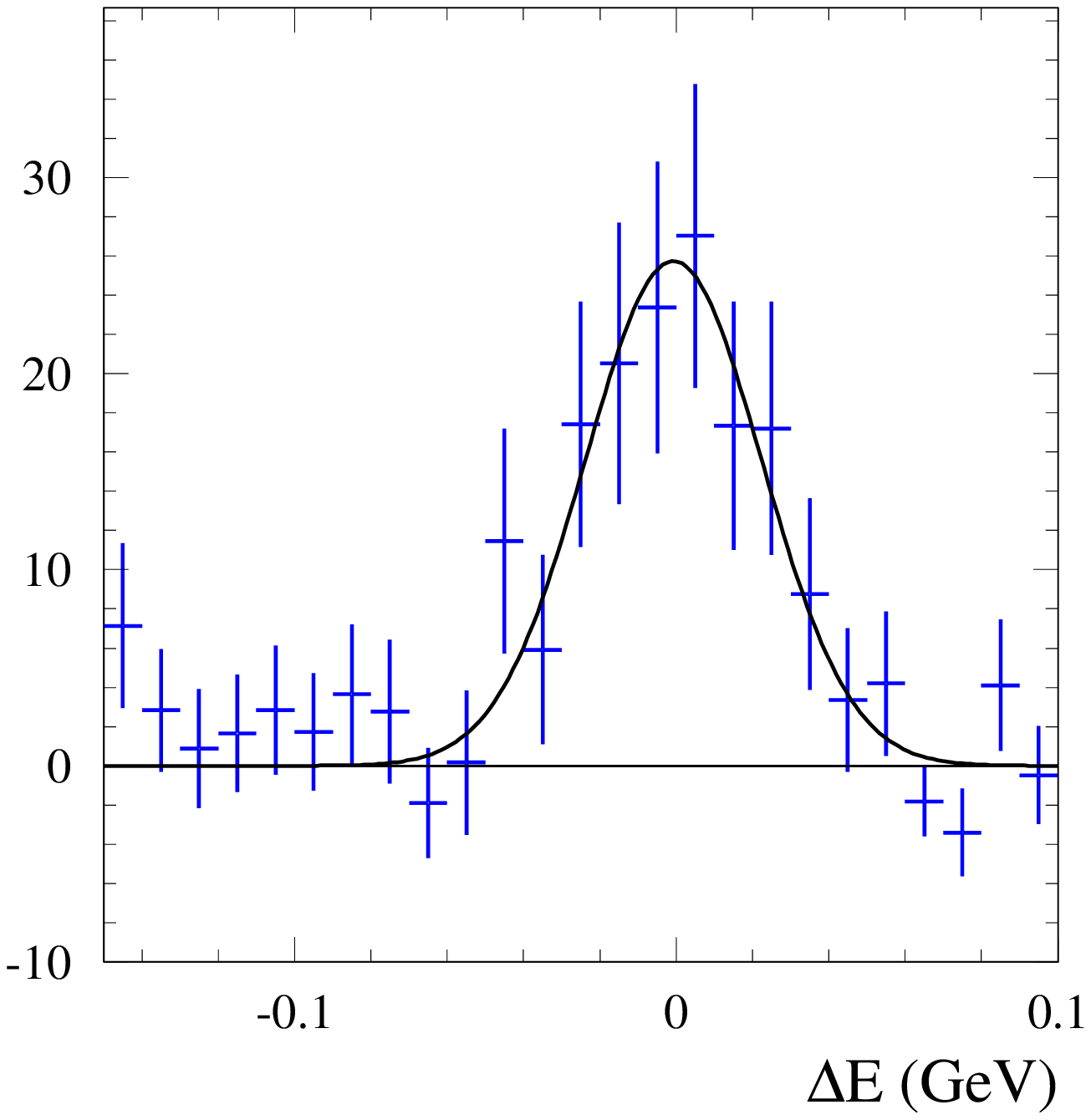,height=1.55in}
\caption{Signal distribution of the $\deltaE$ variable. The left figure is obtained applying a cut on the Likelihood ratio to enrich the data sample in signal events (about 60\% of signal is kept). The right figure shows the $\sPlot$ for signal (all events are kept).}
\label{fig:dE}
\end{figurehere}

Looking at the right plot of Fig.~\ref{fig:dE}, which is a signal $\sPlot$, one can see that these events are signal events, not background events. The pdf of $\deltaE$ which is used in the conventional fit for the whole analysis is superimposed on the $\sPlot$. When this pdf is used, the events in excess are interpreted as background events while performing the fit. Further studies have shown~\cite{bib:TheseMu} that these events are in fact radiative events, \id~$\Bz \to \pipi \gamma$. When ignored in the analysis they lead to underestimates of the branching ratios by about 10\%. The updated results~\cite{bib:hh2005} for the $\Bz \to \pipi$, $\kppim$ analysis, now taking into account the contribution of radiative events, show agreement with the estimate made in Ref.~\cite{bib:TheseMu}.

\section{Applications}
\label{sec:app}
Beside providing a convenient and optimal tool to cross-check the analysis by allowing distributions of control variables to be reconstructed and then compared with expectations, the $\sPlot$ formalism can be applied also to extract physics results, which would otherwise be difficult to obtain. For example, one may be willing to explore some unknown physics involved in the distribution of a variable $x$. Or, one may be interested to correct a particular yield provided by the Likelihood fit from a selection efficiency which is known to depend on a variable $x$, for which the pdf is unknown. Provided one can demonstrate ({\em e.g.} through Monte-Carlo simulations) that the variable $x$ exhibits weak correlation with the discriminating variables $y$.

To be specific, one can take the example of a three body decay analysis of a species, the signal, polluted by background. The signal pdf inside the two-dimensional Dalitz plot is assumed to be not known, because of unknown contributions of resonances, continuum and of interference pattern. Since the $x$-dependence of the selection efficiency $\epsilon(x)$ can be computed without {\it a priori} knowledge of the $x$-distributions, one can build the efficiency corrected two-dimensional $\sPlots$ (cf. Eq.~(\ref{eq:masterequation})):
\begin{equation}
\label{eq:Dalitzcorrected}
{1\over\epsilon(\bar{x})} {N_\nast}~_s\tilde\M_\nast(\bar{x}) \delta x = \sum_{\chiant} {1\over\epsilon(x_e)} \sP_\nast(y_e) ~,
\end{equation}
and compute the efficiency corrected yields:
\begin{equation}
\label{eq:br}
N_\nast^\epsilon= \sum_{e=1}^N {\sP_\nast(y_e)\over\epsilon(x_e)} ~.
\end{equation}
Analyses can then use the $\sPlot$ formalism for validation purposes, but also, using Eq.~(\ref{eq:Dalitzcorrected}) and Eq.~(\ref{eq:br}), to probe for resonance structures and to measure branching ratios~\cite{bib:Dalitz}.

\section{Correlation between variables}
\label{sec:corr}
Correlations between variables, if not trivial, are usually assessed by Monte-Carlo simulations. In case significant correlations are observed, one may still use the $\sPlot$ weight of Eq.~(\ref{eq:weightxnotiny}), but then there is a caveat. The distribution obtained with $\sPlot$ cannot be compared directly with the marginal distribution of $x$. In that case, one must rely on Monte-Carlo simulation, and apply the $\sPlot$ technique to the simulated events, in order to obtain Monte-Carlo $\sPlots$. It is these Monte-Carlo $\sPlots$ which are to be compared to the $\sPlot$ obtained with the real data. Stated differently, the $\sPlot$ can still be applied to compare the behaviour of the data with the Monte-Carlo expected behavior, but it loses its simplicity.

\section{Conclusion}
The technique presented in this paper applies when
\begin{itemize}
\item one examines a data sample originating from different sources of events,
\item a Likelihood fit is performed on the data sample to determine the yields of the sources,
\item this Likelihood uses a set $y$ of discriminating variables,
\item keeping aside a control variable $x$ which is statistically uncorrelated to the set $y$.
\end{itemize}
By building $\sPlots$, one can reconstruct the distributions of the control variable $x$, separately for each source present in the data sample. Although no cut is applied (hence, the $\sPlot$ of a given species represents the whole statistics of this species) the distributions obtained are pure in a statistical sense: they are free from the potential background arising from the other species. The more discriminating the variables~$y$, the clearer the $\sPlot$ is. The technique is straightforward to implement; it is available in the ROOT framework under the class TSPlot\cite{bib:ROOT}. It features several nice properties: both the normalizations and the statistical uncertainties of the $\sPlots$ reflect the fit ouputs.

\end{multicols}

\begin{thebibliography}{99}
\bibitem{bib:sPlot} M.~Pivk and F.R.~Le~Diberder, \NIM A 555, 356-369, 2005 ({\ttfamily physics/0402083}).
\bibitem{bib:TheseMu} M.~Pivk, Th\`ese de l'Universit\'e Paris VII, BABAR-THESIS-03/012 (2003), available (in French) at http://tel.ccsd.cnrs.fr (ID 00002991).
\bibitem{bib:pipi2002} The BABAR Collaboration, \Journal{\em Phys. Rev. Lett.}{89}{281802}{2002}.
\bibitem{bib:hh2005} The BABAR Collaboration, {\ttfamily hep-ex/0508046}.
\bibitem{bib:Dalitz} The BABAR Collaboration, \Journal{\em  Phys. Rev. Lett}{93}{181805}{2004}.
\bibitem{bib:ROOT} http://root.cern.ch/root/htmldoc/TSPlot.html
\end{thebibliography}
\end{document}